\documentclass[prb,twocolumn,amsmath,amssymb]{revtex4}
\usepackage{graphicx}
\usepackage{bm}
\usepackage{float}
\begin{document}

\title{MgB$_{2}$ radio-frequency superconducting quantum interference device prepared by atomic force microscope lithography}

\author{M.~Gregor, T. Plecenik, M. Pra\v{s}\v{c}\'{a}k, R. Mi\v{c}unek, M. Kubinec, V. Ga\v{s}par\'{i}k, M. Grajcar, P. K\'{u}\v{s} and A. Plecenik.}
\affiliation{Department of Experimental Physics,Comenius
University, SK-84248 Bratislava, Slovakia}

\date{\today}

\begin{abstract}

A new method of preparation of radio-frequency superconducting
quantum interference devices on MgB$_{2}$ thin films is presented.
The variable-thickness bridge was prepared by a combination of
optical lithography and of the scratching by an atomic force
microscope. The critical current of the nanobridge was 0.35 $\mu$A
at 4.2 K. Non-contact measurements of the current-phase
characteristics and of the critical current vs. temperature have
been investigated on our structures.

\end{abstract}

\pacs{74.72.-h, 85.25.Dq, 74.50.+r, 85.25.Cp, 74.76.Db }

%--------------------------------------------------------------------

\maketitle
\newpage

Since the discovery of superconductivity in of
MgB$_{2}$,\cite{Nagamatsu01} many techniques have been used to
prepare superconducting weak links, the basic element of
superconducting devices. The relatively high critical temperature
of MgB$_2$, its metallic character, good stability and low
anisotropy could guarantee superior properties of MgB$_2$ weak
links and superconducting quantum interference devices (SQUID).
Weak links with non-tunnel-type conductivity have been attracting
increasing attention, such as point contact, Dayem type bridge,
constant and variable thickness bridge. The main advantage of
non-tunnel-type junctions is their low capacitance. Nanobridges,
with dimensions $L_{eff}$ sufficiently small relative to the
superconducting coherence length $\xi$, have nearly ideal
Josephson behavior. The reported value of the coherence length in MgB$_2$
is 5.2~nm.\cite{Finnemore01} Importantly, also a wider bridge can
show Josephson current-phase relationship, provided that the
width of the bridge is comparable to, or smaller then the
effective London penetration depth $\lambda_{eff}$. Reported
values for the MgB$_2$ bulk penetration depth $\lambda$ vary from
140 to 180~nm.\cite{Brinkman01}  Zhang et al.\cite{Zhang01}
reported an MgB$_2$  point contact SQUID working at 19 K.  Burnell
et al.\cite{Burnell01} realized an MgB$_{2}$ thin film SQUID by
localized ion damage, operated up to 20 K. Several groups prepared
SQUIDs using the focused ion beam and produced nanobridges with
dimensions smaller than the effective London penetration depth
$\lambda_{eff}$.\cite{Brinkman01} Burnell et al.\cite{Burnell02}
reported an MgB$_{2}$ SQUID prepared by FIB operated from 10 K to
24 K, with magnetic flux noise at 20 K as low as 14
$\mu\Phi_0/\sqrt{\mathrm{Hz}}$. Atomic Force Microscope (AFM)
lithography has become a powerful technology, which allows to
create structures with a nanometer scale resolution suitable for
cryogenic application.  Atomic force microscope can pattern
surface in different ways - by local oxidation or by scratching.
Some researchers published technology of local electrochemical
oxidation of the surface of Nb and NbN thin films by voltage
biased tip\cite{Fraucher} and produced Dayem and variable
thickness nanobridges. Others produced nanostructures using
tip-surface mechanical interaction leading to surface plowing or
scratching\cite{Irmer98} and produced a typical a Dayem-type
bridge in an Al film with dimensions $100\times100$~nm$^2$.  We
focus our attention on the preparation of an RF-SQUID
variable-thickness bridge in the MgB$_{2}$ superconductor by AFM
scratching through the photoresist. The main advantage of our
approach is the possibility to prepare nanostructures on the hard
materials through the resist without ion damage, which is produced
in focused ion beam etching. By post-etching with low energy Ar
ions we are able to continuously change the dimensions of the bridge.

\begin{figure}
\centering
\includegraphics[width=7cm]{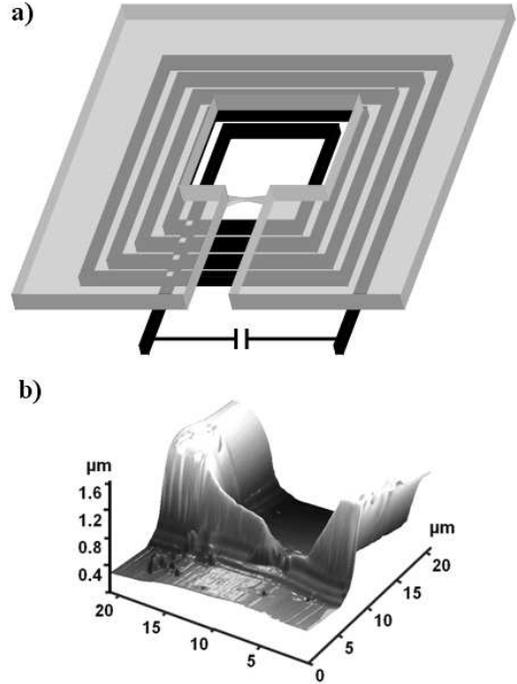}
\caption{ a) A design of the SQUID. b) AFM scan of the MgB$_2$
variable thickness bridge prepared on a 200 nm MgB$_2$ thin film
with a 1.5 $\mu$m thick photoresist} \label{Fig:AFM}
\end{figure}

Superconducting MgB$_2$ thin films were prepared by magnetron
sputtering on the sapphire substrate and ex-situ annealing in a
vacuum chamber. The deposition process was realized by two
independent magnetrons. The technology of MgB$_2$ thin film
preparation was described in Ref. \onlinecite{Micunek06}. Atomic
Force Microscope has been used for surface roughness analysis of
thin films as well as for preparation of the weak links. RMS (Root
Mean Square) roughness of the MgB$_2$ thin films was below 10
nm.\cite{gregor07} The surface scanning has been done in the
semi-contact mode using standard Si tip with W$_2$C coating or
diamond like coating. The maximum lateral scan range of the piezo
tube is $50\times50\ \mu$m$^2$, the maximum vertical range is $2\
\mu$m. Typical curvature radius of the tip is 30~nm and the
resonant frequency is about 255 kHz. The spring constant of the
cantilevers was 20-30 N/m. An RF-SQUID with microstrip (see
Fig.~\ref{Fig:AFM}a) was formed by optical lithography ($3\ \mu$m
wide strip) on the smooth MgB$_2$ thin films. The critical
temperature of a 3-$\mu$m strip was 31.3 K. Our SQUID consists of
a washer ( $50\times50\ \mu$m inner and $2000\times2000\ \mu$m
outer dimension) for better concentration of the magnetic flux and
of a $10\ \mu$m wide slit. We used a positive photoresist for
direct optical lithography and argon ion etching with energy of
the ions 650 eV and ion current 50~mA for determination of the
microstructure. After formation of the structures by Ar ion
etching, the resist was not removed from the MgB$_2$ strip, but
AFM scratching was applied to the micro-strip with $10\ \mu$m
length and $3\ \mu$m width in order to form the variable-thickness
bridge. The AFM scratching (nanolithography) was done
by increasing the deflection of the cantilever in the contact mode.
The width and the depth of the grooves depend on the
deflection force. Because the photoresist is a very soft material
it was hard to determine the sizes of the grooves and the applied force. We
repeatedly scratched through the same place until the photoresist was
removed. By scratching the resist across the strips we produced a
small variable-thickness bridge with submicrometer dimensions. We
applied additional Ar ion beam etching and gradually removed
MgB$_2$ from the strip, so that we achieved the final thickness and width of
the MgB$_2$ strip a few nm and about 100~nm, respectively.
Figure \ref{Fig:AFM}b) shows the variable-thickness bridge
before removing the photoresist (thickness of the MgB$_{2}$ film
was 200 nm and thickness of the photoresist was 1.4 $\mu$m). The
investigated SQUID was placed close to a parallel LC tank circuit.
This resonant detection system consists of a solenoid copper coil
($L_T\approx 73$~nH) and a small capacitance ($C_T\approx700$~pF).
The tank circuit is fed from a current source at the resonant
frequency 24~MHz, which produces alternating flux in the SQUID of
amplitude $MQI_b\sin\omega_0t$, where $Q=80$ is the quality factor
of our tank circuit, $M\approx 0.12$~nH is the mutual inductance
and $I_b\sin\omega_0t$ is the bias current. The DC flux in the
SQUID is produced by the same copper coil. The signal from the
tank circuit is amplified by the first stage of a cold amplifier
located close to the tank circuit at liquid helium temperature.
The cryogenic amplifier consists of a high electron mobility field
effect transistor promoting low noise figure (0.4dB), high gain
(18dB), and wide dynamic range. This amplifier stage provides the
impedance matching to the $50\ \Omega$ coaxial cable and
approximately 10dB gain. The output signal is further amplified by
the next stages at room temperature. The second stage is of a similar
design as the cryogenic stage and the last stage is a 30dB
(Monolithic Microware Integrated Circuit) gain block. The overall
gain is above 50dB with noise figure about 0.5dB at room
temperature. The amplified signal is fed to a phase sensitive
detector (Lock-in amplifier) from which the output is the
amplitude and phase of the signal relating to the reference signal
from the RF generator. Data acquisition and the measurement
process is computer controlled via GPIB interface.
%(See Fig.\ref{schema}).
\begin{figure}[htbp]
\includegraphics[width=7cm]{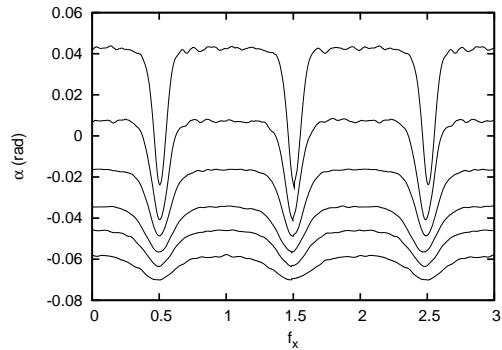}
\caption{Phase shift between the voltage across the tank and the
driving current measured at various temperatures as a function of
the normalized external magnetic flux $f_x=\Phi_x/\Phi_0$. From
top to bottom, the data correspond to $T$=4.2, 6.4, 7.8, 9, 10.4,
and 12.6 K.}
 \label{Fig:alpha}
\end{figure}

\begin{figure}
\includegraphics[width=7cm]{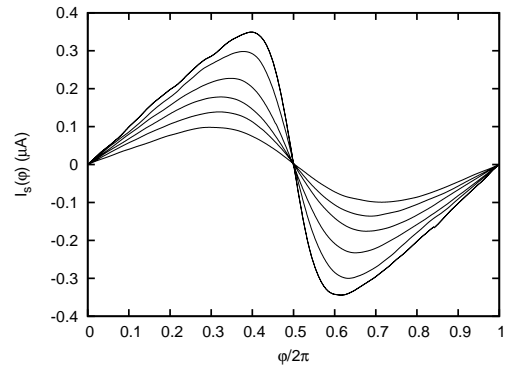}
\caption{$I_s(\varphi)$  characteristics calculated from the phase
characteristics (see Fig.~\ref{Fig:alpha}). The curves corresponds
to temperature, from top to bottom, $T$=4.2, 6.4, 7.8, 9, 10.4,
and 12.6 K.} \label{Fig:Iphi}
\end{figure}

\begin{figure}
\includegraphics[width=7cm]{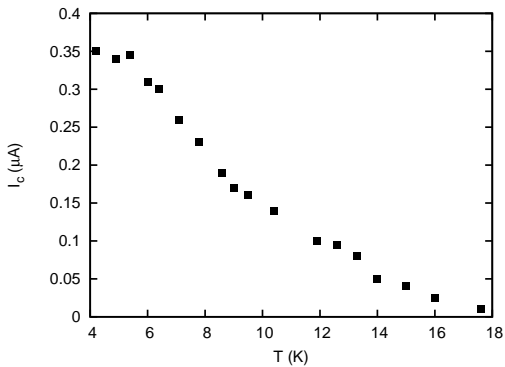}
\caption{Temperature dependence of the critical current of the
weak link.} \label{Fig:IcT}
\end{figure}

The SQUID phase characteristics $\alpha(f_x)$, where $\alpha$ is
the phase shift between the bias current $I_b$ and the voltage
across the tank circuit, and $f_x=\Phi_e/\Phi_0$ is external magnetic flux
$\Phi_e$ normalized to the magnetic flux quantum $\Phi_0$,
are shown in Fig.\ref{Fig:alpha}. In order
to obtain more information on the microbridge superconducting weak
link, we have calculated its current-phase relationship (CPR)
$I_s(\varphi)=I_cf(\varphi)$ using the modified Rifkin-Deaver
method\cite{Rifkin76, Golubov04}. The resonance frequency of the
tank circuit depends on the effective inductance defined
as\cite{Silver67,Barone}
\begin{equation}
    L_{eff}(\varphi)=L\left(1+\frac{1}{\beta f'(\varphi)}\right)\ ,
    \label{Eq:Lj}
\end{equation}
where L is the inductance of the washer, $\varphi=2\pi\Phi/\Phi_0$ is
the gauge-invariant phase difference across the Josephson
junction, $\Phi$ is the internal flux in the washer, $\Phi_0$ is
the elementary magnetic flux quantum and $\beta=2\pi LI_c/\Phi_0$
is the normalized SQUID inductance. If $\beta<1$, \textit{i.e.} if
the SQUID is in a nonhysteretic mode,  we can determine the
current-phase relationship from the relation\cite{Golubov04}
\begin{equation}
    I_s(\varphi)=\frac{L_T(\Delta I)^2}{2\pi Q
    \Phi_0}\int_0^\varphi\tan\alpha(\phi)d\phi\ ,
    \label{Eq:Ip}
\end{equation}
where $L_T$ is inductance of the tank coil, $\Delta I=16.5~\mu A$
is the magnitude of the current in the coil producing magnetic
flux $\Phi_0$ in the rf squid, $Q$ is the quality factor of the
tank circuit and $\phi=2\pi \Phi_e/\Phi_0$ is the normalized
external \textit{dc} magnetic flux. The phase shift $\alpha(\phi)$
between the voltage across the tank and the driving current is
measured experimentally. The calculated current-phase
characteristics are shown in Fig.~\ref{Fig:Iphi}. The maximum of
the $I_s(\varphi)$ characteristics is shifted towards
$\varphi=\pi$. Such deviation from standard  Josephson sinusoidal
behavior was predicted by Kulik and
Omelyanchuk\cite{Kulik75,Kulik77} for short Josephson junctions
$d_{eff}\ll\xi$, where $d_{eff}$ is the effective length of the
weak link.\cite{Likharev79} However, this condition can be hardly
satisfied since $\xi\approx 5$~nm. Thus, we believe that the weak
links are in the dirty limit and the shift of the maximum of the
$I_s(\varphi)$ characteristics is caused by the enhanced kinetic
inductance of the junction. The kinetic inductance can be
characterized by the parameter\cite{Likharev79}
\begin{equation}
    l=2\pi\frac{{\cal L}I_c}{\Phi_0}\ ,
    \label{Eq:l}
\end{equation}
where ${\cal L}$ is the kinetic inductance of the weak links. If
$l$ is larger than unity the $I_s(\varphi)$ becomes  multivalued
and the SQUID is in a hysteretic regime even for $\beta<1$. This
parameter is important since the large value of $l$ deteriorates
parameters of the superconducting weak links, especially the SQUID
performance.\cite{Likharev79} However, our SQUID is in the
nonhysteretic regime, i.e. with $l<1$ as one can expect for
variable-thickness bridges.\cite{Likharev79,Jackel76} The
thickness, width and length of the bridge in the thinnest region
is estimated from the AFM picture (Fig.~\ref{Fig:AFM}b) to be
$~10$, 200, and 100~nm, respectively.

Also the temperature dependence (compare Fig.~\ref{Fig:IcT} with
Fig.~9 in Ref.~\onlinecite{Likharev79} or Fig.~14 in
Ref.~\onlinecite{Golubov04}) indicates that the weak link is long
with $d_{eff}>\xi$ and that the critical temperature of the bridge
$T'_c$ is suppressed ($T'_c<T_c$). Similar temperature dependences
were measured also by other groups. For example, AFM
(Fig.~\ref{Fig:IcT}), focused-ion-beam patterned
nanobridges\cite{Brinkman01}, as well as MgB$_{2}$ break
junctions\cite{gonnelli} have similar behavior of the critical
current as a function of temperature with critical temperature,
$T_c\approx 20$~K. It is striking that different technologies of
weak link preparation give nearly identical results.

To conclude, our results show that AFM scratching through a soft
material (in our case the photoresist) is a powerful technology
which can produce MgB$_2$ weak links with a single valued Josephson
current-phase relationship ($l<1$). Even if etching was performed
step-by-step without the contact for checking the etching
speed, reproducibility of this process was 80\%. Such weak links
are promising for practical application and they can be
incorporated into Josephson effect
devices.\cite{Likharev79,Likharev}

\acknowledgments

This work was supported by grant VEGA 1/2011/05, APVT projects No.
APVT-51-016604, APVT-20-011804 and project aAV/1126/2004.

%----------------------------------------------------------------
{\bf Refferences}

%----------------------------------------------------------------

\end{document}